\documentclass[11pt,a4paper]{article}
\usepackage[truedimen,margin=33mm]{geometry}

\usepackage{mathrsfs}
\usepackage{amssymb}
\usepackage{amsmath}
\usepackage{ascmac}
\usepackage{amsthm}
\usepackage[dvipdfmx]{graphicx}

\usepackage{color}

\usepackage{titlesec}
\titleformat*{\section}{\large\bfseries}
\titleformat*{\subsection}{\it}

\newtheorem{thm}{Theorem}

\newtheorem{as}{Assumption}

\usepackage{setspace}

\def\barx{{\bar{\x}}}

\def\bary{{\bar{y}}}

\def\ep{{\varepsilon}}
\def\la{{\lambda}}
\def\si{{\sigma}}

\def\th{{\theta}}

\def\bbe{{\text{\boldmath $\beta$}}}

\def\bphi{{\text{\boldmath $\phi$}}}

\def\phih{{\widehat \phi}}
\def\sih{{\widehat \si}}

\def\tah{{\widehat \tau}}
\def\muh{{\widehat \mu}}
\def\lah{{\widehat \la}}

\def\mut{{\widetilde \mu}}

\def\bbeh{{\widehat \bbe}}

\def\bphih{{\widehat \bphi}}

\def\Si{{\Sigma}}
\def\bSi{{\text{\boldmath $\Si$}}}
\def\bPhi{{\text{\boldmath $\Phi$}}}

\def\0{{\text{\boldmath $0$}}}
\def\1{{\text{\boldmath $1$}}}
\def\b{{\text{\boldmath $b$}}}

\def\x{{\text{\boldmath $x$}}}
\def\y{{\text{\boldmath $y$}}}
\def\z{{\text{\boldmath $z$}}}

\def\I{{\text{\boldmath $I$}}}

\def\X{{\text{\boldmath $X$}}}

\def\tr{{\rm tr\,}}

\def\E{{\rm E}}
\def\P{{\rm P}}
\def\Cov{{\rm Cov}}
\def\Cor{{\rm Cor}}
\def\Re{\mathbb{R}}

\def\mitt{\vspace{-0.15cm}\item}

\title{{\bf Adaptively Transformed Mixed Model Prediction of General Finite Population Parameters}}

\date{}

\begin{document}

\maketitle
\doublespacing

\begin{center}
\vspace{-1.7cm}\noindent
SHONOSUKE SUGASAWA\\
{\it Center for Spatial Information Science, The University of Tokyo}\\
TATSUYA KUBOKAWA\\
{\it Faculty of Economics, The University of Tokyo}\\
\end{center}

\vspace{0.3cm}\noindent
{\bf Abstract.}\ \ 
For estimating area-specific parameters (quantities) in a finite population, a mixed model prediction approach is attractive.
However, this approach strongly depends on the normality assumption of the response values although we often encounter a non-normal case in practice.
In such a case, transforming observations to make them suitable for normality assumption is a useful tool, but the problem of selecting suitable transformation still remains open.
To overcome the difficulty, we here propose a new empirical best predicting method by using a parametric family of transformations to estimate a suitable transformation based on the data.
We suggest a simple estimating method for transformation parameters based on the profile likelihood function, which achieves consistency under some conditions on transformation functions.
For measuring variability of point prediction, we construct an empirical Bayes confidence interval of the population parameter of interest.
Through simulation studies, we investigate numerical performance of the proposed methods.
Finally, we apply the proposed method to synthetic income data in Spanish provinces in which the resulting estimates indicate that the commonly used log-transformation would not be appropriate.


\newpage
\section{Introduction}\label{sec:intro}

The mixed model prediction based on random effect models has been widely used in small area estimation (Rao and Molina, 2015).
The random effect models used in small area estimation are mainly divided into two models: the Fay-Herriot model (Fay and Herriot, 1979) and the nested error regression model (Battese et al., 1988).
Especially, the nested error regression model has been used for estimating population parameters in a finite population.
Here we consider a finite population consisting of $m$ areas and each area has $N_i$ units for $i=1,\ldots,m$.
Let $Y_{ij}$ be a characteristic of the $j$th unit in the $i$th area, the main purpose is to estimate the area-specific parameter defined as 
\begin{equation}\label{mu}
\mu_i=\frac1{N_i}\sum_{j=1}^{N_i}T(Y_{ij}),
\end{equation}
where $T(\cdot)$ is a known (user-specified) function.
The simplest choice is $T(x)=x$, in which $\mu_i$ corresponds to the finite population mean, and  this case has been studied in the literature; Chambers et al. (2014), Jiang and Lahiri (2006), Lahiri and Mukherjee (2007) and Schmit et al. (2016).
On the other hand, as noted by Molina and Rao (2010), more complex forms of $T(\cdot)$ are often used in practice.
For example, in poverty mapping, we often adopt the FGT poverty measure $T(x)=\{(z-x)/z\}^{\alpha}I(x<z)$ (Foster et al., 1984), where $z$ is a suitable poverty threshold.

If all the units $Y_{ij}$ in the $i$th area were observed, we could calculate the true value of $\mu_i$.
However, only a part of the units are available in practice.
Let $n_i(<N_i)$ be the number of sampled units and $y_s=\{y_{ij},\ j=1,\dots,n_i, \ i=1,\ldots,m\}$ be the sampled data.
It is known that the direct estimator of $\mu_i$ using the observed units has high variability, especially in the case that $n_i$ is much smaller than $N_i$.
In real application, some covariates associated with $Y_{ij}$ are available not only for sampled but also for non-sampled units, which are denoted by $\x_{ij}$ with $j=1,\ldots,N_i$ and $i=1,\ldots,m$.
Hence, one aims to estimate $\mu_i$ based on the sampled data and information on covariates.
To this end, a typical strategy is to assume that all the units follow the nested error regression model (Battese et al., 1988): 
\begin{equation}\label{NER}
Y_{ij}=\x_{ij}^t\bbe+v_i+\ep_{ij}, \ \ \ j=1,\ldots,N_i, \ \ i=1,\ldots,m,
\end{equation}
where $\x_{ij}$ and $\bbe$ are $p$-dimensional vectors of covariates and regression coefficients, $v_i$ is the area-specific effect which follows $N(0,\tau^2)$ and $\ep_{ij}$ is a sampling error distributed as $N(0,\si^2)$.
Note that the model (\ref{NER}) leads to normality assumption of $Y_{ij}$.
Then, the conditional distribution of the non-sampled data $Y_{ij}$ given all the sampled data $y_s$ is given by
\begin{equation}\label{pos}
Y_{ij}| y_s\sim N\bigg(\x_{ij}^t\bbe+\frac{n_i\tau^2}{\si^2+n_i\tau^2}(\bary_i-\barx_i^t\bbe), \ \frac{\si^2\tau^2}{\si^2+n_i\tau^2}\bigg), \ \ \ j=n_i+1,\ldots,N_i,
\end{equation}
which follows from the normality of $Y_{ij}$ under the model (\ref{NER}).
Then the best predictor of $\mu_i$ under squared error loss is the conditional expectation $\E[\mu_i|y_s]$, which has the form
\begin{equation}\label{BP}
\mut_i\equiv\E[\mu_i|y_s]=\frac1{N_i}\bigg\{\sum_{j=1}^{n_i}T(y_{ij})+\sum_{j=n_i+1}^{N_i}\E[T(Y_{ij})|y_s]\bigg\}.
\end{equation}
Here, the expectation $\E[T(Y_{ij})|y_s]$ could be computed via the Monte Carlo integration by generating a large number of random samples from the conditional distribution (\ref{pos}).
Moreover, the best predictor $\mut_i$ depends on the unknown model parameters $\bbe,\tau^2$ and $\si^2$ in the model (\ref{NER}), so that these parameters should be replaced with their estimated counterparts.
To this end, the model parameters in the model (\ref{NER}) would be estimated based on the sampled data $y_s$ by using, for example, the maximum likelihood or restricted maximum likelihood methods.

It is observed that the key assumption in deriving the best predictor (\ref{BP}) is the normality of $Y_{ij}$ in the model (\ref{NER}), which enables us to obtain the simple expression of the conditional distribution (\ref{pos}).
However, we often encounter the case where the normality assumption is not plausible for $Y_{ij}$.
For instance, in poverty mapping, $Y_{ij}$ would be a welfare variable (e.g. income), so that the distribution of $Y_{ij}$ would be right skewed and would not be normal.
In this case, several methods have been considered so far. 
For instance, Chambers and Tzavidis (2006) proposed the M-quantile models and Sinha and Rao (2009) and Chambers et al. (2014) among others proposed robust methods against outliers under normality assumptions.
However, these methods basically aims to estimate only the mean parameters, corresponding $T(x)=x$ in (\ref{mu}), and do not take account of the data characteristics like skewness.
On the other hand, Molina and Rao (2010) considered the nested error regression model (\ref{NER}) for the transformed variables $H(Y_{ij})$ instead of $Y_{ij}$, which would be able to take account of data characteristics.
For instance, the log-transformation, $H(x)=\log x$, would be the most standard approach in small area estimation (e.g. Slud and Haiti, 2006; Molina and Martin, 2017) when $Y_{ij}$ is right skewed as often appeared in welfare variables.
However, we can not know how skew the true data distribution is whereas the use of log-transformation forces the amount of skewness in the data distribution.
This problem can be regarded as the misspecification of the transformation, under which the predictor of $\mu_i$ might be biased or inefficient.
In such a case, it would be more preferable to consider a parametric family of transformations and we propose to estimate the transformation based on the data for solving the misspecification as much as possible.
This approach would enable us to get better prediction although the problem of violating normality assumption are not necessarily solved due to limitations arising from the use of a parametric family of transformations.
We derive a form of the best predictor of $\mu_i$ and provide a simple estimating method for transformation parameters based on profile likelihood function, which produces a consistent estimator under some regularity conditions.
We also construct an empirical Bayes confidence interval of $\mu_i$ for measuring the variability of the point prediction. 
The proposed intervals are shown to have $O(m^{-1})$ coverage error, and we also suggest the parametric bootstrap calibration for confidence intervals with further accuracy.

This paper is organized as follows: In Section \ref{sec:TP}, we describe the proposed prediction method as well as parameter estimation of the model parameters.
In Section \ref{sec:EBCI}, we construct an empirical Bayes confidence interval of $\mu_i$.
In Section \ref{sec:num}, we present the results from simulation studies and a data application.
In Section \ref{sec:conc},  we give conclusions and some discussions.
The technical proofs are given in Appendix.

\section{Adaptively Transformed Mixed Model Prediction}\label{sec:TP}


\subsection{Transformed best predictor}
Let $H_{\la}(\cdot)$ be a family of transformations with parameter $\la$.
The transformation parameter $\la$ might be multidimensional, but we treat $\la$ as a scalar parameter for notational simplicity.
The assumptions and specific choices of $H_{\la}(\cdot)$ will be discussed in the subsequent section.
We assume that the transformed variable $H_{\la}(y_{ij})$ follows the nested error regression model:
\begin{equation}\label{TNER}
H_{\la}(Y_{ij})=\x_{ij}^t\bbe+v_i+\ep_{ij}, \ \ \ j=1, \ldots, N_i, \ \ \ i=1,\ldots,m,
\end{equation}
where $\x_{ij}$ and $\bbe$ are $p$-dimensional vectors of fixed covariates and regression coefficients, $v_i$ and $\ep_{ij}$ are an area-specific effect and a sampling error, respectively.
Here we assume that $v_i$ and $\ep_{ij}$ are mutually independent and distributed as $v_i\sim N(0,\tau^2)$ and $\ep_{ij}\sim N(0,\si^2)$ with unknown two variance parameters $\tau^2$ and $\si^2$.
It is worth noting that, owing to the area effect $v_i$, the units in the same area are mutually correlated while the units in the different area are independent.
Specifically, from (\ref{TNER}), it holds $\Cor(H_{\la}(Y_{ij}),H_{\la}(Y_{ik}))=(\tau^2+\si^2)^{-1}\tau^2, \ j\neq k$, thereby the units in the same area are mutually correlated and the degree of correlation is determined by the ratio $\tau^2/\si^2$.
The normality assumptions of $v_i$ and $\ep_{ij}$ leads to the assumption that $H_{\la}(Y_{ij})\sim N(\x_{ij}^t\bbe,\tau^2+\si^2)$.
Although, there would not necessarily exist the suitable transformation parameter to make the transformed response hold the normality assumption exactly, our theory will be developed under the normality assumption.

In model (\ref{TNER}), the response variable $Y_{ij}$ is assumed to follow a class of distributions defined by transforming the normal distribution $N(\x_{ij}^t\bbe,\tau^2+\si^2)$ by $H_{\la}^{-1}(\cdot)$, the inverse function of $H_{\la}(\cdot)$.
For example, the use of log-transformation leads to the assumption that $Y_{ij}$ follows a log-normal distribution.
Moreover, the use of a transformation does not change the correlation structure that the units in the same areas are correlated while different areas are kept independent.
On the other hand, the mean and variance $Y_{ij}$ could be a complicate function of the mean $\x_{ij}^t\bbe$ and variance $\tau^2+\si^2$ in the transformed scale, so that it could induce some relationships between mean and variance as observed in a log-normal distribution, and the correlations among $Y_{ij}$'s in the same areas would not be necessarily constant.

Let $y_s=\{y_{ij},\ j=1,\dots,n_i, \ i=1,\ldots,m\}$ be the sampled data.
From the model (\ref{TNER}), we have $H_{\la}(Y_{ij})|y_s\sim N(\th_{ij},s_i^2+\si^2)$, $j=n_i+1,\ldots, N_i$, where
\begin{equation}\label{tht}
\th_{ij}=\x_{ij}^t\bbe+\frac{\tau^2}{\si^2+n_i\tau^2}\sum_{j=1}^{n_i}(H_{\la}(y_{ij})-\x_{ij}^t\bbe), \ \ \ 
s_i=\sqrt{\frac{\si^2\tau^2}{\si^2+n_i\tau^2}}.
\end{equation}
Hence, the best predictor of $\mu_i$ given in (\ref{mu}) can be obtained as
\begin{equation}\label{TBP}
\mut_i(y_s)\equiv\E[\mu_i|y_s]
=\frac1{N_i}\left\{\sum_{j=1}^{n_i}T(y_{ij})+\sum_{j=n_i+1}^{N_i}\E[T\circ H_{\la}^{-1}(u_{ij})]\right\},
\end{equation}
where the expectation is taken with respect to $u_{ij}\sim N(\th_{ij}, s_i^2+\si^2)$, and $T\circ H_{\la}^{-1}(\cdot)$ is the composite function of $T(\cdot)$ and $H_{\la}^{-1}$, the inverse function of $H_{\la}(\cdot)$.
Although the expectation $\E[T\circ H_{\la}^{-1}(u_{ij})]$ does not have a closed form in general, it can be easily computed via the Monte Carlo integration.
We call the best predictor (\ref{TBP}) adaptively transformed best predictor (ATP).


\subsection{Estimation of structural parameters}
Let $\bphi=(\bbe^t,\tau^2,\si^2,\la)^t$ be a vector of unknown model parameters in (\ref{TNER}).
We here propose estimating $\bphi$ based on the following log-marginal likelihood function:
\begin{equation}\label{ML}
\begin{split}
L(\bphi)&=-\frac12\sum_{i=1}^m\log|\bSi_i|-\frac12\sum_{i=1}^m\left\{H_{\la}(y_i)-\X_i\bbe\right\}^t\bSi_i^{-1}\left\{H_{\la}(y_i)-\X_i\bbe\right\}\\
&  \ \ \ \ \ \  -\frac{1}2\sum_{i=1}^mn_i\log 2\pi+\sum_{i=1}^m\sum_{j=1}^{n_i}\log H'_{\la}(y_{ij}),
\end{split}
\end{equation}
where $(\bSi_i)_{k\ell}=\tau^2+\si^2I(k=\ell)$, $H_{\la}(y_i)=(H_{\la}(y_{i1}),\ldots,H_{\la}(y_{in_i}))^t$, $\X_i=(\x_{i1}^t,\ldots,\x_{in_i}^t)^t$, and $H'_{\la}(\cdot)$ denotes the derivative of $H_{\la}(\cdot)$.
The maximum likelihood estimator of $\bphi$ can be defined as the maximizer of $L(\bphi)$.

For maximizing the likelihood function $L(\bphi)$, we first note that the profile log-likelihood function of $\la$ can be expressed as
\begin{equation}\label{PL}
\text{PL}(\la)=\text{ML}(\la)+\sum_{i=1}^m\sum_{j=1}^{n_i}\log H'_{\la}(y_{ij}),
\end{equation}
where $\text{ML}(\la)$ is the maximum log-likelihood of the nested error regression model with response values $H_{\la}(y_{ij})$ and covariate vectors $\x_{ij}$.
Since the computation of $\text{ML}(\la)$ can be readily carried out via well-developed numerical methods, e.g. \verb+sae+ package in R (Molina and Marhuenda, 2015), the point evaluation of the profile likelihood (\ref{PL}) is quite easy.
Hence, we may obtain the maximizer of $\text{PL}(\la)$ by using, for example, the golden section method (Brent et al., 1973).
Once we obtain the estimator $\lah$, we get the estimators of other parameters by applying the nested error regression model to the data set $\{H_{\lah}(y_{ij}),\x_{ij}\}$.

For estimating the two variance parameters $\tau^2$ and $\si^2$, the restricted maximum likelihood (REML) method (Jiang, 1996) might be more attractive than the maximum likelihood method in terms of estimating variance components since the REML method is known to produce estimates with smaller bias than the maximum likelihood method.  
To implement the REML estimation, the first three terms in (\ref{ML}) are changed to those of the REML method, and the resulting function could be maximized by profiling as used in the maximum likelihood method.
We note that the estimator of transformation parameters would be changed by adapting the REML estimation, and it would be hard to investigate theoretical differences in terms of estimating transformation parameters. 
Therefore, we consider only the maximum likelihood estimator for simplicity.


\subsection{Class of transformations}\label{sec:trans}
When applying the aforementioned prediction method, the concrete choice of $H_{\la}(\cdot)$ would be important in practice.
For the choice of suitable transformations, we need to take account of the data characteristics.
For instance, income data is often right skewed, so that a family of transformations that includes the log-transformation would be suitable.

Before considering the concrete family of transformations, we introduce the following class of transformations for theoretical guarantee of the proposed method.

\vspace{0.2cm}
\begin{as}\label{as:trans}
(Class of transformations)
\begin{itemize}
\item[1.]
$H_{\la}$ is a differentiable and monotone function, and the range of $H_{\la}$ is $\Re$ for all $\la$.

\mitt[2.]
For fixed $x$, $H_{\la}(x)$ as the function of $\la$ is differentiable.

\mitt[3.]
The function $|\partial H_{\la}(w)/\partial\la|$, $|\partial^2 H_{\la}(w)/\partial \la^2|$ and $|\partial^2 \log H'_{\la}(w)/\partial\la^2|$ with $w=H_{\la}^{-1}(x)$ are bounded from the upper by $C_1\{\exp(C_2x)+\exp(-C_2x)\}$ with some constants $C_1,C_2>0$.
\end{itemize}
\end{as}

\vspace{0.5cm}
The first condition is crucial in this context.
If the range of $H_{\la}$ is not $\Re$, but some subset $A\subset\Re$, the inverse function $H_{\la}^{-1}$ cannot be defined on $\Re\setminus A$, which causes problems in computing the predictor (\ref{TBP}).
Although the Box-Cox transformation (Box and Cox, 1964) is widely used for positive valued data and has been adopted in small area estimation (Li and Lahiri, 2007), it does not belong to the class of transformations due to the limited range of the Box-Cox transformation.

When we focus on estimating poverty indicators based on welfare variables like income, the following two properties of parametric transformations would be preferable: one is simplicity of the inverse function $H^{-1}_{\la}(x)$ used in computing the predictor (\ref{TBP}), and the other is eliminating skewness of data distribution which would be often the case in income distribution.
There area several parametric transformations related to the Box-Cox transformation, e.g. John and Draper (1980) and Yeo and Johnson (2000).
However, the transformation by John and Draper (1980) cannot necessarily eliminate the skewness since the transformation is an odd function.
On the other hand, the transformation by Yeo and Johnson (2000) has a relatively complicated form and so does the inverse transformation.
As transformations that belongs to the class and satisfies two desirable properties, we consider two parametric transformations, the dual power (DP) transformation (Yang, 2006) and the sinh-arcsinh (SS) transformation (Jones and Pewsey, 2009).

The DP transformation is described as 
\begin{equation}\label{DPT}
H_{\la}^{\rm DP}(x)=\frac{x^{\la}-x^{-\la}}{2\la}, \ \ \ \  x>0, \ \ \ \la>0,
\end{equation}
where $\lim_{\la\to 0}H_{\la}^{\rm DP}(x)=\log x$.
In the context of small area estimation, the Fay-Herriot model has been extended by Sugasawa and Kubokawa (2017) with use of the DP transformation. 
It is easy to confirm that the range of DP transformation is $\Re$.
Note that $\lambda$ controls the skewness of the transformed random variable $H_{\la}^{\rm DP}(y_{ij})$.
The inverse function and the Jacobian appeared in the predictor (\ref{TBP}) and the profile likelihood (\ref{PL}), respectively, are given by 
$$
H_{\la}^{\text{DP}(-1)}(x)=\Big(\la x+\sqrt{1+\la^2x^2}\Big)^{1/\la}\ \ \ \ \text{and} \ \ \ \ 
\frac{d H_{\la}^{\rm DP}(x)}{dx}=\frac12(x^{\la-1}+x^{-\la-1}).
$$
The original DP transformation (\ref{DPT}) by Yang (2006) can be applied only for a positive valued response. 
However, as demonstrated in Section \ref{sec:pov}, welfare variables can take negative values depending on their definition.
In this case, we propose the shifted-DP (SDP) transformation of the form $H_{\la,c}(x)=\{(x+c)^{\la}-(x+c)^{-\la}\}/2\la$, where $c\in (\min(y_{ij})+\ep,\infty)$ with some small $\ep>0$.

The SS transformation has the following form:
\begin{equation}\label{SS}
H_{a,b}^{\rm SS}(x)=\sinh(b\sinh^{-1}(x)-a),\ \ \ x\in (-\infty,\infty), \ \ \ a\in (-\infty,\infty), \ \ \ b\in (0,\infty),
\end{equation}
where $\sinh(x)=(e^x-e^{-x})/2$ is the hyperbolic sine function, $\sinh^{-1}(x)=\log(x+\sqrt{x^2+1})$, and two transformation parameter $a$ and $b$ control skewness and tail heaviness of the transformed random variable $H_{a,b}^{\rm SS}(y_{ij})$, respectively.  
The inverse transformation and the Jacobian are obtained as
$$
H_{a,b}^{\text{SS}(-1)}(x)=\sinh(b^{-1}\sinh^{-1}(x)+a), \ \ \ \  \text{and}\ \ \ \ \ 
\frac{dH_{a,b}^{\rm SS}(x)}{dx}=b\sqrt{\frac{1+H_{a,b}^{\rm SS}(x)^2}{1+x^2}}.
$$
Since the domain of the SS transformation is the whole real line, it could be used not only for positive valued data but also real valued data.

As mentioned at the beginning of this section, which transformation should be used would depend on the data characteristics.
Meanwhile, we may select the suitable parametric transformation in a more objective way by using information criteria of the form: $-2\text{ML}+a(N)(p+q+2)$, where ML is the maximum log-likelihood, namely the maximum value of (\ref{ML}), $q$ is the number of transformation parameters and $N=\sum_{i=1}^mn_i$ is the total number of sampled units.
Setting $a(N)=2$ and $a(N)=\log N$ correspond to AIC-like and BIC-like criterion, respectively.


\subsection{Large sample properties}
We here consider the large sample properties of the estimator of structural parameters.
To this end, we assume the following condition:
\begin{as}\label{as:model}
(Assumptions under large $m$)
\begin{itemize}
\item[1.]
The true parameter vector $\bphi_0$ is an interior point of the parameter space $\bPhi$.

\mitt[2.]
$0<\min_{i=1,\ldots,m}N_i\leq \max_{i=1,\ldots,m}N_i<\infty$.

\mitt[3.]
The elements of $\X_i$ are uniformly bounded and $\X_i^t\X_i$ is positive definite. 

\mitt[4.]
$m^{-1}\sum_{i=1}^m\X_i^t\bSi_i^{-1}\X_i$ converges to a positive definite matrix as $m\to\infty$.
\end{itemize}
\end{as}

The first condition implies that there exists the true transformation parameter that the transformed random variable $H_{\la}(Y_{ij})$ achieves normality while it might not necessarily hold in practice. 
Since the asymptotic variance and covariance matrix of the maximum likelihood estimator can be derived from the Fisher information matrix, we first provide the Fisher information matrix in the following theorem, where the proof is given in Appendix.

\begin{thm}\label{thm:asymp}
When the Fisher information is denoted by $I_{\phi_k\phi_j}=-\E[\partial^2 L(\bphi)/\partial\phi_k\partial\phi_j]$, we have 
\begin{align*}
&
I_{\tau^2\tau^2}=\frac12\sum_{i=1}^m(\1_{n_i}^t\bSi_i^{-1}\1_{n_i})^2, \ \ \ \ 
I_{\tau^2\si^2}=\frac12\sum_{i=1}^m\1_{n_i}^t\bSi_i^{-2}\1_{n_i}, \ \ \ \ 
I_{\si^2\si^2}=\frac12\sum_{i=1}^m\tr(\bSi_i^{-2}),\\
&
\I_{\bbe\bbe}=\sum_{i=1}^m\X_i^t\bSi_i^{-1}\X_i, \ \ \ \ 
\I_{\bbe\tau^2}=\I_{\bbe\si^2}=0, \ \ \ \ 
I_{\la\si^2}=-\sum_{i=1}^m\E\left[\z_i^t\bSi_i^{-2}H_{\la}^{(1)}(y_i)\right],\\
&
\I_{\la\bbe}=-\sum_{i=1}^m\X_i^t\bSi_i^{-1}\E\left[H_{\la}^{(1)}(y_i)\right], \ \ \ \ 
I_{\la\tau^2}=-\sum_{i=1}^m\E\left[\z_i^t\bSi_i^{-1}\1_{n_i}\1_{n_i}^t\bSi_i^{-1}H_{\la}^{(1)}(y_i)\right],\\
&
I_{\la\la}=\sum_{i=1}^m\E\left[H_{\la}^{(1)}(y_i)^t\bSi_i^{-1}H_{\la}^{(1)}(y_i)\right]+\sum_{i=1}^m\E\left[\z_i^t\bSi_i^{-1}H_{\la}^{(2)}(y_i)\right]-\sum_{i=1}^m\sum_{j=1}^{n_i}\E\left[\frac{\partial^2}{\partial\la^2}\log H_{\la}'(y_{ij})\right],
\end{align*}
where $H_{\la}^{(k)}(y_i)=\partial^k H_{\la}(y_i)/\partial\la^k$ for $k=1,2$, $\z_i=H_{\la}(y_i)-\X_i\bbe$, and $\E[\cdot]$ denotes the expectation with respect to $y_{ij}$'s following the model (\ref{TNER}).
Then, under Assumptions \ref{as:trans} and \ref{as:model}, the maximum likelihood estimator $\bphih$ is asymptotically distributed as $\bphih\sim N(\bphi,\I_{\bphi}^{-1})$.
\end{thm}

From Theorem \ref{thm:asymp}, it is observed that the information matrix of $(\bbe^t,\tau^2,\si^2)$ does not depend on the transformation parameter $\la$, and their expressions are the same as those of the traditional nested error regression models.
While the two variance parameters $\tau^2$ and $\si^2$ are orthogonal to $\bbe$ in the sense that $\I_{\bbe\tau^2}=\I_{\bbe\si^2}=0$, the transformation parameter $\la$ is not orthogonal to the others.
The expectations appeared in the Fisher matrix is not analytically tractable, but it can be easily estimated by replacing the expectation with its sample counterpart.
In the case that $\la$ is multidimensional, the extension of Theorem \ref{thm:asymp} is straightforward. 
The expressions of $H_{\la}^{(k)}(y_i)$ and $\partial^2 \log H_{\la}'(y_{ij})/\partial\la^2$ could be analytically complicated and require tedious algebraic calculations.
In such a case, the numerical derivative would be useful since we need to compute only the point values of the derivatives.

\section{Empirical Bayes Confidence Intervals}\label{sec:EBCI}

\subsection{Asymptotically valid confidence intervals}\label{sec:NCI}
Measuring the variability of the transformed empirical best predictor $\muh_i$ is an important issue in practice. 
Traditionally, the mean squared error (MSE) of $\muh_i$ has been used, and several methods ranging from analytical method (Prasad and Rao, 1990) to numerical methods (Hall and Maiti, 2006) have been considered.
On the other hand, an empirical Bayes confidence interval of $\mu_i$ is more preferable since it can provide distributional information than MSE although construction of the confidence interval is generally difficult.
Here, we derive an asymptotically valid empirical Bayes confidence interval of $\mu_i$.

The key to derivation of the confidence interval is the conditional distribution of $\mu_i$ given $y_i$.
Noting that $\Cov(H_{\la}(Y_{ij}),H_{\la}(Y_{ik})|y_i)=\text{Var}(v_i|y_i)=s_i^2$ for $j\neq k$, it follows that
$$
(H_{\la}(Y_{i,n_i+1}),\ldots,H_{\la}(Y_{iN_i}))^t|y_i\sim N((\th_{i,n_i+1},\ldots,\th_{iN_i})^t,s_i^2\1_{N_i-n_i}\1_{N_i-n_i}^t+\si^2\I_{N_i-n_i}),
$$
namely, the each component has the expression
$$
H_{\la}(Y_{ij})|y_i=\th_{ij}+s_iz_i+\si w_{ij}, \ \ \ j=n_i+1,\ldots,N_i,
$$
where $z_i$ and $w_{ij}$ are mutually independent standard normal random variables, and $\th_{ij}$ and $s_i$ are defined in (\ref{tht}).
Then the posterior distribution of $\mu_i$ can be expressed as 
\begin{equation}\label{pos-mu}
\mu_i|y_i\stackrel{\text{d}}{=}
\frac1{N_i}\left\{\sum_{j=1}^{n_i}T(y_{ij})+\sum_{j=n_i+1}^{N_i}T\circ H_{\la}^{-1}\left(\th_{ij}+s_iz_i+\si w_{ij}\right)\right\},
\end{equation}
which is a complex function of standard normal random variables $z_i$ and $w_{ij}$.
However, random samples from the conditional distribution (\ref{pos-mu}) can be easily simulated.

We define $Q_{a}(y_i,\bphi)$ as the lower $100a\%$ quantile point of the posterior distribution of $\mu_i$ with the true $\bphi$, which satisfies $\P(\mu_i\leq Q_{a}(y_i,\bphi)|y_i)=a$.
Hence, the Bayes confidence interval of $\mu_i$ with nominal level $1-\alpha$ is obtained as $I_{\alpha}=( Q_{\alpha/2}(y_i,\bphi), Q_{1-\alpha/2}(y_i,\bphi))$, which holds that
$\P(\mu_i\in I_{\alpha})=1-\alpha$.
However, the interval $I_{\alpha}$ depends on the unknown parameter $\bphi$, so that the feasible version of $I_{\alpha}$ is obtained by replacing $\bphi$ with its estimator $\bphih$, namely
\begin{equation}\label{naive}
I_{\alpha}^{N}=( Q_{\alpha/2}(y_i,\bphih), Q_{1-\alpha/2}(y_i,\bphih)),
\end{equation}
which we call naive empirical Bayes confidence interval of $\mu_i$.
The two quantiles appeared in (\ref{naive}) can be computed by generating a large number of random samples from the conditional distribution (\ref{pos-mu}).
Owing to the asymptotic properties of $\bphih$, the coverage probability of the naive interval (\ref{naive}) converges to the nominal level as the number of areas $m$ tends to infinity as shown in the following theorem proved in Appendix.

\begin{thm}\label{thm:cp1}
Under Assumptions \ref{as:trans} and \ref{as:model}, it holds $\P(\mu_i\in I_{\alpha}^{N})=1-\alpha+O(m^{-1})$.
\end{thm}


\subsection{Bootstrap calibrated intervals}\label{sec:BCI}

As shown in Theorem \ref{thm:cp1}, the coverage error of the naive interval (\ref{naive}) is of order $m^{-1}$, which is not necessarily negligible when $m$ is not sufficiently large.
Since the number of $m$ is usually moderate in practice, the calibrated intervals with higher accuracy would be valuable.
Following Chatterjee, et al. (2008), Hall and Maiti (2006), we construct a second order corrected empirical Bayes confidence interval $I_{\alpha}^{C}$ satisfying $P(\mu_i\in I_{\alpha}^{C})=1-\alpha+o(m^{-1})$.

To begin with, we define the bootstrap estimator of the coverage probability of the naive interval.
Let $Y_{ij}^{\ast}$ be the parametric bootstrap samples generated from the estimated model (\ref{TNER}) with $\bphi=\bphih$, and $y_i^{\ast}=\{Y_{ij}^{\ast},  \ j=1,\ldots,n_i\}$.
Moreover, let $\mu_i^{\ast}$ be the bootstrap version of $\mu_i$ based on $Y_{ij}^{\ast}$'s.  
Since the coverage probability is $\P( Q_{a/2}(y_i,\bphih)\leq \mu_i\leq Q_{1-a/2}(y_i,\bphih))$, its parametric bootstrap estimator can be defined as
$$
\text{CP}(a)=\E^{\ast}\left[I\left\{Q_{a/2}(y_i^{\ast},\bphih^{\ast})\leq \mu_i^{\ast}\leq Q_{1-a/2}(y_i^{\ast},\bphih^{\ast})\right\}\right],
$$
where the expectation is taken with respect to the bootstrap samples $Y_{ij}^{\ast}$'s.
Based on the coverage probability, we define the calibrated nominal level $a^{\ast}$ as the solution of the equation $\text{CP}(a^{\ast})=1-\alpha$, which can be solved by the bisectional method (Brent, 1973).
Then, the calibrated interval is given by
\begin{equation}\label{bootCI}
I_{\alpha}^{C}=( Q_{a^{\ast}/2}(y_i,\bphih), Q_{1-a^{\ast}/2}(y_i,\bphih)),
\end{equation}
which has second order accuracy as shown in the following theorem proved in Appendix.

\begin{thm}\label{thm:cp2}
Under Assumptions \ref{as:trans} and \ref{as:model}, it holds $\P(\mu_i\in I_{\alpha}^{C})=1-\alpha+o(m^{-1})$.
\end{thm}

\section{Numerical Studies}\label{sec:num}


\subsection{Evaluation of prediction errors}
We first evaluate the prediction errors of the proposed predictors together with some existing methods. 
To this end, we considered the following data generating processes: 
\begin{equation*}
\begin{split}
&\text{(A)}\  \ 
(2\la)^{-1}(Y_{ij}^{\la}-Y_{ij}^{-\la})=\mu_{ij}+v_i+\ep_{ij}, \ \ \ v_i\sim N(0,\tau^2), \ \ \ \ep_{ij}\sim N(0,\si^2)\\
&\text{(B)}\ \ 
(2\la)^{-1}(Y_{ij}^{\la}-Y_{ij}^{-\la})=\mu_{ij}+v_i+\ep_{ij}, \ \ \ v_i\sim t_5(0,\tau^2), \ \ \ \ep_{ij}\sim t_5(0,\si^2)\\
&\text{(C)}\ \ 
Y_{ij}=\exp(\mu_{ij})v_i\ep_{ij}, \ \ \ v_i\sim \Gamma(1/\tau^2,1/\tau^2), \ \ \ \ep_{ij}\sim \Gamma(1/\si^2,1/\si^2)\\
&\text{(D)}\  \ 
Y_{ij}=0.2\exp(U_{ij})+0.8U_{ij}^2, \ \ \ 
U_{ij}=\mu_{ij}+v_i+\ep_{ij}, \\
& \ \ \ \ \ \ \ \ \ \ \ \ v_i\sim N(0,\tau^2), \ \ \ \ep_{ij}\sim N(0,\si^2),
\end{split}
\end{equation*}
where $i=1,\ldots,m$, $j=1,\ldots,N$, $\mu_{ij}=\beta_0+\beta_1x_{1ij}+\beta_2x_{2ij}+\beta_3x_{3ij}$, $(\beta_0,\beta_1,\beta_2,\beta_3)=(2,1,-0.5,1)$, $\tau=0.5$, $\si=0.8$, and $x_{1ij}, x_{2ij}$ and $x_{3ij}$ were generated from Bernoulli distributions with probabilities $0.3, 0.5$ and $0.5$, respectively.
Based on the above models, we considered seven scenarios of data generating processes as summarized in Table \ref{tab:sim-scenario}.
In this study, we focus on estimating the following parameters:
\begin{equation}\label{pr}
\mu_i=\frac1N\sum_{j=1}^NT_{\alpha}(Y_{ij}), \ \ \ \ \ \ \
T_{\alpha}(x)=\left(\frac{z-x}{z}\right)^{\alpha}I(x<z),
\end{equation}
where $z$ is defined as $0.6$ times median of $Y_{ij}$'s.
Note that $T_{\alpha}(x)$ is known as FGT poverty measures (Foster et al., 1984).
We considered two cases of $\alpha$, $\alpha=0$ (poverty rate) and $\alpha=1$ (poverty gap).
We set $m=30$ and divided $m$ areas into five groups with equal number of areas, and we set the same numbers of sampled units $n_i$ within the same groups.
The group pattern of $n_i$ was $(10,20,30,40,50)$.
For the size of units $N$, we considered two cases, $N=200$ and $N=400$, to check sensitivity of ratios $n_i/N$.

Among the generated $Y_{i1},\ldots,Y_{iN}$, we used first $n_i$ observations $y_{i1}(=Y_{i1}),\ldots,y_{in_i}(=Y_{in_i})$ as the sampled data.
Then, based on the sampled data $y_{ij}$'s and covariates $X_{ij}=(x_{1ij},x_{2ij},x_{3ij})^t$, we predict $\mu_i$ based on the proposed adaptively transformed empirical best prediction (ATP) method with DP transformation (\ref{DPT}) and the transformed empirical best prediction (TP) (Molina and Rao, 2010) with log-transformation. 
We used 1000 Monte Carlo samples in applying these two methods.
As a competitor from other model-based methods, we employed the M-quantile method (Chambers and Tzavidis, 2006).
We fitted the M-quantile model to the sampled data in the same way as in Chambers and Tzavidis (2006), and modified the distribution function estimator given in equation (5) in Chambers and Tzavidis (2006) to an estimator of $\mu_i$ by replacing the indicator function with $T_{\alpha}(\cdot)$. 
Moreover, we computed the following direct estimator (DE): $\muh_i^{D}=n_i^{-1}\sum_{j=1}^{n_i}T_{\alpha}(y_{ij})$. 
Note that the proposed model is correctly specified in Scenarios (s1)$\sim$(s3), namely, there exists the true transformation parameter such that the transformed variable achieves normality.
On the other hand, in Scenarios (s4)$\sim$(s7), the proposed model is misspecified in the sense that there is no true transformation parameter that achieves normality

To compare the performances of the four methods, we computed mean squared error (MSE) defined as 
$$
\text{MSE}_i=\frac1R\sum_{r=1}^R\Big(\muh_i^{(r)}-\mu_i^{(r)}\Big)^2,
$$ 
with $R=500$, where $\muh_i^{(r)}$ and $\mu_i^{(r)}$ are the estimated and true values of $\mu_i$, respectively, in the $r$th iteration.
The obtained values of MSEs are averaged within the same groups and the results are reported in Tables \ref{tab:comp1} and \ref{tab:comp2}.
From these tables, we can observe that the proposed method provides better estimates than the three existing methods in almost all cases, and there are not much differences between the two cases of $N$.
In scenario (s1), the performance between ATP and TP are almost the same while the ATP method is overfitting while the log-transformed model is correctly specified.
Meanwhile, in the other scenarios, the ATP method can improve the estimation accuracy of TP method as well as MQ and DE methods, by adaptively estimating the transformation parameter from the data even when the model assumption in the TP method is violated.
In Supplementary Material, we provide additional simulation results (e.g. relative bias and coefficient of variations).

\begin{table}[!htb]
\centering
\caption{
8 Scenarios of simulation studies. 
}
\label{tab:sim-scenario}

\medskip
\begin{tabular}{cccccccccc}
\hline
Scenario & s1 & s2 & s3 & s4 & s5 & s6 & s7  \\
\hline
Model & A & A & A  & B & B & C & D \\
$\lambda$ & $0$ & $0.25$ & $0.50$  & $0.25$ & $0.50$ & - & - \\ 
\hline
\end{tabular}
\end{table}

\begin{table}[!htb]
\vspace{1cm}
\centering
\caption{
The group-wise averaged values of mean squared errors (MSE) of poverty rate $(\alpha=0)$ for four methods (ATP, LTP, DE and MQ) in 7 scenarios.
All the values in the table are multiplied by $1000$.
}
\label{tab:comp1}

\medskip
\begin{tabular}{ccccccccccccccc}
\hline
 &  & & \multicolumn{4}{c}{$N=200$} && \multicolumn{4}{c}{$N=400$}\\
 Scenario & $n$  &  & ATP & LTP & MQ & DE &  & ATP & LTP & MQ & DE\\
\hline
 & 10 &  & 5.04 & 5.04 & 15.08 & 21.44 &  & 5.00 & 5.02 & 16.55 & 17.51\\
 & 20 &  & 3.33 & 3.32 & 12.48 & 9.14 &  & 3.34 & 3.33 & 14.09 & 12.01\\
s1 & 30 &  & 2.60 & 2.60 & 11.65 & 7.08 &  & 2.34 & 2.33 & 14.32 & 8.14\\
 & 40 &  & 1.76 & 1.76 & 9.76 & 4.62 &  & 1.80 & 1.79 & 11.44 & 5.05\\
 & 50 &  & 1.51 & 1.51 & 10.41 & 3.40 &  & 1.37 & 1.37 & 12.39 & 3.28\\
 \hline
 & 10 &  & 5.45 & 5.78 & 13.90 & 21.28 &  & 5.10 & 5.37 & 14.87 & 17.49\\
 & 20 &  & 3.09 & 3.35 & 10.62 & 8.32 &  & 3.19 & 3.43 & 11.31 & 11.70\\
s2 & 30 &  & 2.35 & 2.58 & 9.75 & 6.30 &  & 2.34 & 2.53 & 11.74 & 8.12\\
 & 40 &  & 1.73 & 1.87 & 8.64 & 4.35 &  & 1.79 & 2.00 & 9.84 & 5.02\\
 & 50 &  & 1.42 & 1.62 & 8.52 & 3.37 &  & 1.43 & 1.66 & 10.68 & 3.31\\
 \hline
 & 10 &  & 4.62 & 5.95 & 11.37 & 18.12 &  & 4.43 & 5.83 & 12.27 & 15.71\\
 & 20 &  & 2.90 & 3.68 & 9.85 & 7.84 &  & 2.91 & 4.05 & 9.92 & 10.77\\
s3 & 30 &  & 2.13 & 3.02 & 8.10 & 5.88 &  & 2.01 & 3.02 & 10.03 & 6.85\\
 & 40 &  & 1.56 & 2.19 & 7.05 & 4.13 &  & 1.57 & 2.48 & 8.08 & 4.38\\
 & 50 &  & 1.31 & 2.18 & 7.00 & 3.07 &  & 1.30 & 2.27 & 8.43 & 3.09\\
 \hline
 & 10 &  & 6.03 & 6.47 & 13.72 & 21.22 &  & 5.28 & 5.61 & 13.92 & 16.41\\
 & 20 &  & 3.39 & 3.65 & 10.83 & 8.26 &  & 3.45 & 3.77 & 10.74 & 10.98\\
s4 & 30 &  & 2.47 & 2.69 & 10.03 & 6.09 &  & 2.43 & 2.71 & 11.04 & 8.02\\
 & 40 &  & 1.78 & 2.00 & 8.15 & 4.49 &  & 1.79 & 2.07 & 9.72 & 4.57\\
 & 50 &  & 1.53 & 1.74 & 8.76 & 3.30 &  & 1.53 & 1.84 & 9.97 & 3.25\\
 \hline
 & 10 &  & 5.16 & 6.82 & 11.39 & 18.59 &  & 4.73 & 6.63 & 10.48 & 14.41\\
 & 20 &  & 3.17 & 4.42 & 8.59 & 7.66 &  & 3.10 & 4.60 & 8.94 & 10.25\\
s5 & 30 &  & 2.11 & 3.26 & 7.39 & 5.54 &  & 2.14 & 3.49 & 8.72 & 6.76\\
 & 40 &  & 1.52 & 2.50 & 6.21 & 3.93 &  & 1.76 & 2.94 & 7.47 & 4.52\\
 & 50 &  & 1.41 & 2.47 & 6.31 & 3.06 &  & 1.37 & 2.61 & 7.78 & 2.90\\
 \hline
 & 10 &  & 6.20 & 6.69 & 17.21 & 21.35 &  & 6.23 & 6.83 & 18.06 & 18.38\\
 & 20 &  & 3.88 & 4.28 & 13.73 & 9.16 &  & 3.86 & 4.29 & 14.48 & 12.47\\
s6 & 30 &  & 2.76 & 3.11 & 11.96 & 6.90 &  & 2.92 & 3.38 & 14.75 & 8.93\\
 & 40 &  & 2.07 & 2.33 & 10.42 & 4.62 &  & 2.10 & 2.42 & 12.16 & 5.11\\
 & 50 &  & 1.68 & 1.96 & 10.56 & 3.52 &  & 1.69 & 2.03 & 13.02 & 3.52\\
 \hline
 & 10 &  & 5.74 & 8.45 & 13.91 & 19.94 &  & 6.03 & 8.99 & 14.63 & 17.19\\
 & 20 &  & 3.83 & 5.82 & 11.71 & 8.80 &  & 3.89 & 6.48 & 12.19 & 11.83\\
s7 & 30 &  & 2.82 & 4.82 & 9.55 & 6.46 &  & 2.93 & 5.47 & 11.00 & 8.38\\
 & 40 &  & 2.19 & 3.95 & 8.23 & 4.48 &  & 2.30 & 4.53 & 9.27 & 4.99\\
 & 50 &  & 2.06 & 4.02 & 8.77 & 3.55 &  & 1.98 & 4.27 & 10.46 & 3.55\\
\hline
\end{tabular}
\vspace{1cm}
\end{table}

\begin{table}[!htb]
\vspace{1cm}
\centering
\caption{
The group-wise averaged values of mean squared errors (MSE) of poverty rate $(\alpha=1)$ for four methods (ATP, LTP, DE and MQ) in 7 scenarios.
All the values in the table are multiplied by $1000$.
}
\label{tab:comp2}

\medskip
\begin{tabular}{ccccccccccccccc}
\hline
 &  & & \multicolumn{4}{c}{$N=200$} && \multicolumn{4}{c}{$N=400$}\\
 Scenario & $n$  &  & ATP & LTP & MQ & DE &  & ATP & LTP & MQ & DE\\
 \hline
 & 10 &  & 1.86 & 1.86 & 4.16 & 6.52 &  & 1.83 & 1.83 & 4.36 & 4.76\\
 & 20 &  & 1.17 & 1.17 & 3.25 & 2.27 &  & 1.22 & 1.21 & 3.59 & 3.63\\
s1 & 30 &  & 0.92 & 0.92 & 3.14 & 2.14 &  & 0.83 & 0.83 & 3.84 & 2.39\\
 & 40 &  & 0.62 & 0.62 & 2.36 & 1.43 &  & 0.66 & 0.65 & 2.96 & 1.56\\
 & 50 &  & 0.52 & 0.52 & 2.59 & 1.06 &  & 0.50 & 0.50 & 3.12 & 1.05\\
 \hline
 & 10 &  & 1.76 & 1.80 & 4.05 & 5.58 &  & 1.73 & 1.77 & 4.12 & 4.41\\
 & 20 &  & 0.94 & 0.96 & 2.81 & 2.04 &  & 1.02 & 1.03 & 3.11 & 3.02\\
s2 & 30 &  & 0.77 & 0.80 & 2.79 & 1.79 &  & 0.74 & 0.76 & 3.28 & 2.10\\
 & 40 &  & 0.55 & 0.56 & 2.28 & 1.26 &  & 0.57 & 0.58 & 2.57 & 1.39\\
 & 50 &  & 0.44 & 0.45 & 2.19 & 0.98 &  & 0.45 & 0.46 & 2.80 & 0.91\\
 \hline
 & 10 &  & 1.20 & 1.27 & 3.26 & 3.95 &  & 1.20 & 1.31 & 3.39 & 3.36\\
 & 20 &  & 0.73 & 0.80 & 2.64 & 1.53 &  & 0.77 & 0.83 & 2.90 & 2.33\\
s3 & 30 &  & 0.55 & 0.59 & 2.47 & 1.26 &  & 0.51 & 0.56 & 2.80 & 1.47\\
 & 40 &  & 0.40 & 0.44 & 1.98 & 0.98 &  & 0.43 & 0.48 & 2.45 & 1.01\\
 & 50 &  & 0.34 & 0.37 & 2.00 & 0.78 &  & 0.33 & 0.36 & 2.46 & 0.68\\
 \hline
 & 10 &  & 1.92 & 1.96 & 3.49 & 5.67 &  & 1.67 & 1.70 & 3.49 & 4.05\\
 & 20 &  & 0.97 & 0.97 & 2.45 & 1.93 &  & 1.03 & 1.02 & 2.42 & 2.82\\
s4 & 30 &  & 0.76 & 0.76 & 2.41 & 1.67 &  & 0.73 & 0.73 & 2.63 & 1.93\\
 & 40 &  & 0.55 & 0.54 & 1.81 & 1.23 &  & 0.59 & 0.59 & 2.25 & 1.31\\
 & 50 &  & 0.46 & 0.46 & 1.95 & 0.92 &  & 0.46 & 0.46 & 2.33 & 0.90\\
 \hline
 & 10 &  & 1.28 & 1.39 & 2.97 & 4.08 &  & 1.19 & 1.26 & 2.67 & 2.98\\
 & 20 &  & 0.74 & 0.76 & 2.13 & 1.50 &  & 0.77 & 0.82 & 2.40 & 2.35\\
s5 & 30 &  & 0.51 & 0.54 & 1.93 & 1.21 &  & 0.54 & 0.54 & 2.30 & 1.45\\
 & 40 &  & 0.39 & 0.40 & 1.54 & 0.90 &  & 0.45 & 0.47 & 2.06 & 1.03\\
 & 50 &  & 0.36 & 0.38 & 1.62 & 0.75 &  & 0.35 & 0.37 & 2.13 & 0.68\\
 \hline
 & 10 &  & 2.61 & 2.64 & 6.77 & 7.63 &  & 2.65 & 2.73 & 6.81 & 6.66\\
 & 20 &  & 1.57 & 1.58 & 5.11 & 3.30 &  & 1.59 & 1.59 & 5.37 & 4.44\\
s6 & 30 &  & 1.10 & 1.12 & 4.78 & 2.52 &  & 1.24 & 1.26 & 5.67 & 3.07\\
 & 40 &  & 0.84 & 0.84 & 3.98 & 1.68 &  & 0.95 & 0.94 & 4.70 & 1.89\\
 & 50 &  & 0.69 & 0.70 & 3.93 & 1.35 &  & 0.74 & 0.74 & 5.03 & 1.32\\
 \hline
 & 10 &  & 2.45 & 2.67 & 5.47 & 7.32 &  & 2.57 & 2.82 & 5.55 & 5.99\\
 & 20 &  & 1.54 & 1.67 & 4.20 & 2.86 &  & 1.66 & 1.86 & 4.71 & 4.42\\
s7 & 30 &  & 1.12 & 1.26 & 3.82 & 2.37 &  & 1.19 & 1.36 & 4.30 & 3.02\\
 & 40 &  & 0.88 & 0.97 & 2.99 & 1.74 &  & 0.97 & 1.09 & 3.48 & 1.87\\
 & 50 &  & 0.77 & 0.90 & 3.06 & 1.37 &  & 0.78 & 0.92 & 3.89 & 1.20\\
 \hline
\end{tabular}
\vspace{1cm}
\end{table}


\subsection{Performance of empirical Bayes confidence intervals}
We next evaluate the finite sample performance of the empirical Bayes confidence intervals given in Section \ref{sec:EBCI}.
To this end, we adopted the same settings as used in scenario (s2) in Section 4.1 and focused on the same population parameters $\mu_i$ given in (\ref{pr}) with $\alpha=0$ and $1$. 
For constructing confidence intervals of $\mu_i$, we employed two methods: bootstrap calibrated confidence interval (\ref{bootCI}) as well as the naive confidence interval (\ref{naive}), which are denoted by BCI and NCI, respectively.
We used 200 Monte Carlo samples and 100 bootstrap replications in applying these two methods.
Note that theoretical coverage accuracy of BCI and NCI is $o(m^{-1})$ and $O(m^{-1})$, where $m=30$ in this study.

To evaluate the performances of the two confidence intervals, based on $R=500$ simulation runs, we computed the following empirical coverage probability (CP) and average length of confidence interval (AL): 
$$
{\rm CP}_i=\frac1R\sum_{r=1}^RI(\mu_i^{(r)}\in\text{CI}_i^{(r)})  \ \ \ \ \text{and} \ \ \ \ 
{\rm AL}_i=\frac1R\sum_{r=1}^R|\text{CI}_i^{(r)}|,
$$
where $\mu_i^{(r)}$ and $\text{CI}_i^{(r)}$ are the true value and the confidence interval in the $r$th iteration.
We averaged CP's and AL's within the same groups, and reported the results in Table \ref{tab:CI}. 
Table \ref{tab:CI} shows that NCI tends to produce shorter confidence intervals and the coverage probability is smaller than the nominal level for all areas.
On the other hand, BCI produces more accurate confidence intervals than NCI, which would support the theoretical property given in Theorem \ref{thm:cp2}.
Since underestimation of risk estimates may yield serious problems in practice, BCI would be appealing when the number of areas is not large.

\begin{table}[!htb]
\centering
\caption{
The group-wise averaged values of coverage probability (CP) and average length (AL) of $95\%$ confidence intervals based on the naive method (NCI) and the bootstrap method (BCI).}
\label{tab:CI}

\medskip
\begin{tabular}{ccccccccccccccc}
\hline
 & $(N,\alpha)$ && \multicolumn{2}{c}{$(200,0)$} & \multicolumn{2}{c}{$(200,1)$} && \multicolumn{2}{c}{$(400,0)$} & \multicolumn{2}{c}{$(400,1)$}\\
 & $n$ &  & BCI & NCI & BCI & NCI && BCI & NCI & BCI & NCI\\
 \hline
 & 10 &  & 0.958 & 0.944 & 0.951 & 0.936 && 0.949 & 0.938 & 0.951 & 0.935\\
 & 20 &  & 0.946 & 0.930 & 0.944 & 0.933 && 0.952 & 0.932 & 0.946 & 0.925\\
CP & 30 &  & 0.951 & 0.935 & 0.952 & 0.936 && 0.951 & 0.941 & 0.952 & 0.936\\
 & 40 &  & 0.947 & 0.933 & 0.956 & 0.936 && 0.953 & 0.942 & 0.954 & 0.939\\
 & 50 &  & 0.951 & 0.934 & 0.954 & 0.936 && 0.955 & 0.942 & 0.957 & 0.942\\
 \hline
 & 10 &  & 0.281 & 0.264 & 0.155 & 0.145 && 0.272 & 0.256 & 0.150 & 0.140\\
 & 20 &  & 0.218 & 0.205 & 0.118 & 0.111 && 0.221 & 0.208 & 0.123 & 0.115\\
AL & 30 &  & 0.187 & 0.175 & 0.101 & 0.095 && 0.185 & 0.174 & 0.102 & 0.096\\
 & 40 &  & 0.166 & 0.155 & 0.088 & 0.083 && 0.164 & 0.154 & 0.090 & 0.084\\
 & 50 &  & 0.151 & 0.142 & 0.081 & 0.076 && 0.145 & 0.136 & 0.079 & 0.074\\
 \hline
\end{tabular}
\end{table}


\subsection{Example: poverty mapping in Spain}\label{sec:pov}
We applied the proposed method to estimation of poverty indicators in Spanish provinces, using the synthetic income data available in \verb+sae+ package (Molina and Marhuenda, 2015) in \verb+R+ language.
The similar data set was used in Molina and Rao (2010) and Molina et al. (2014). 
Such data are available for $m=52$ areas and the sample sizes (the number of observed units) range from $20$ to $1420$, so that there are no out-of-sample areas.
The total number of samples units is $17199$.
The welfare variable for the individuals is the equivalized annual net income denoted by $E_{ij}$, noting that the small portions of $E_{ij}$ take negative values.
The median of $E_{ij}$ is about $10800$ and area-wise medians of $E_{ij}$ range about from $7500$ to $13700$, so that the area-wise distributions would be quite different.
As auxiliary variables, we considered the indicators of the four groupings of ages (16-24, 25-49, 50-64 and $\geq$65), the indicator of having Spanish nationality, the indicators of education levels (primary education and post-secondary education), and the indicators of two employment categories (employed, unemployed).
An intercept term is also included in our model.

Let $\x_{ij}$ be the vector of auxiliary variables including an intercept term.
We consider the four models with different families of transformations, SDP transformation, SDP transformation with known shift (SDP-s), SS transformation and shifted log-transformation (SL), which are described as
\begin{equation}\label{pv-model}
\begin{split}
\text{SDP:} \ \ \ \ 
&(2\la)^{-1}\left\{(E_{ij}+c)^{\la}-(E_{ij}+c)^{-\la}\right\}=\x_{ij}^t\bbe+v_i+\ep_{ij},\\
\text{SDP-s:} \ \ \ \ 
&(2\la)^{-1}\left\{(E_{ij}+c^{\ast})^{\la}-(E_{ij}+c^{\ast})^{-\la}\right\}=\x_{ij}^t\bbe+v_i+\ep_{ij},\\
\text{SS:} \ \ \ \ 
&\sinh(b\sinh^{-1}(E_{ij})-a)=\x_{ij}^t\bbe+v_i+\ep_{ij},\\
\text{SL:} \ \ \ \ 
&\log(E_{ij}+c^{\ast})=\x_{ij}^t\bbe+v_i+\ep_{ij},\\
\end{split}
\end{equation}
where $v_i\sim N(0,\tau^2)$, $\ep_{ij}\sim N(0,\si^2)$ and $c^{\ast}=|\min(E_{ij})|+1$.
By maximizing the profile likelihood function of transformation parameters, we obtained the following estimates:
\begin{align*}
&\text{(SDP)} \ \ \lah=0.090\ (1.99\times 10^{-3}), \ \ \ \ \widehat{c}=4319 \ (170.69)\\
&\text{(SDP-s)} \ \ \lah=0.290 \ (8.18\times 10^{-4})\\
&\text{(SS)} \ \widehat{a}=-0.584 \ (6.82\times 10^{-2}) , \ \ \ \ \widehat{b}=0.463 \ (5.62\times 10^{-3}),
\end{align*}
where the values in the parentheses are the corresponding standard errors calculated from the Fisher information matrix given in Theorem \ref{thm:asymp}.
From the above result, it can be observed that the approximate $95\%$ confidence intervals of the transformation parameter $\la$ in SDP and SDP-s are bounded from $0$, which means that the log-transformed model would be inappropriate.

\begin{figure}[!htb]
\centering
\includegraphics[width=15cm,clip]{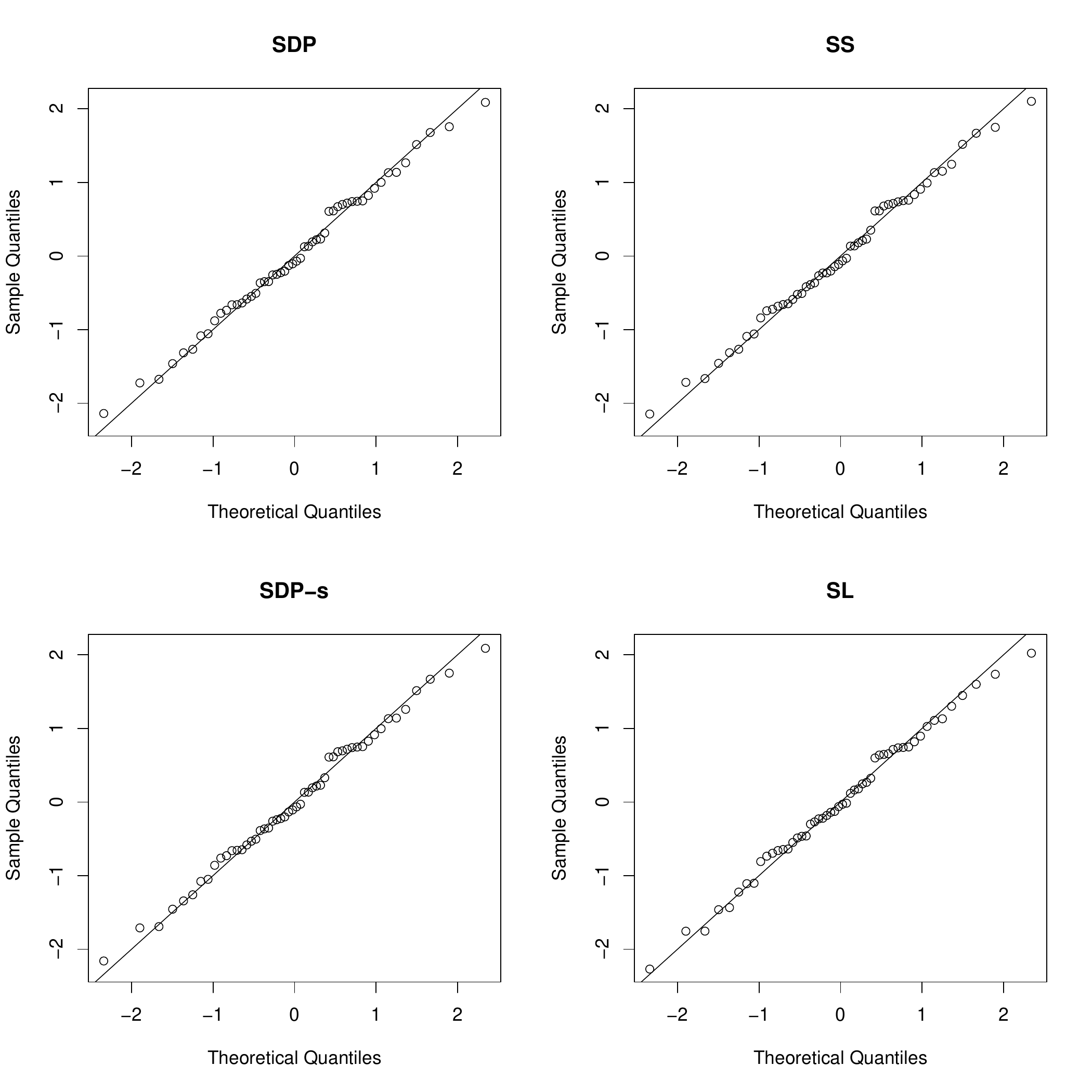}
\caption{QQ-plots of standardized random effect estimates in the four models.}
\label{fig:resid1}
\end{figure}

\begin{figure}[!htb]
\centering
\includegraphics[width=15cm,clip]{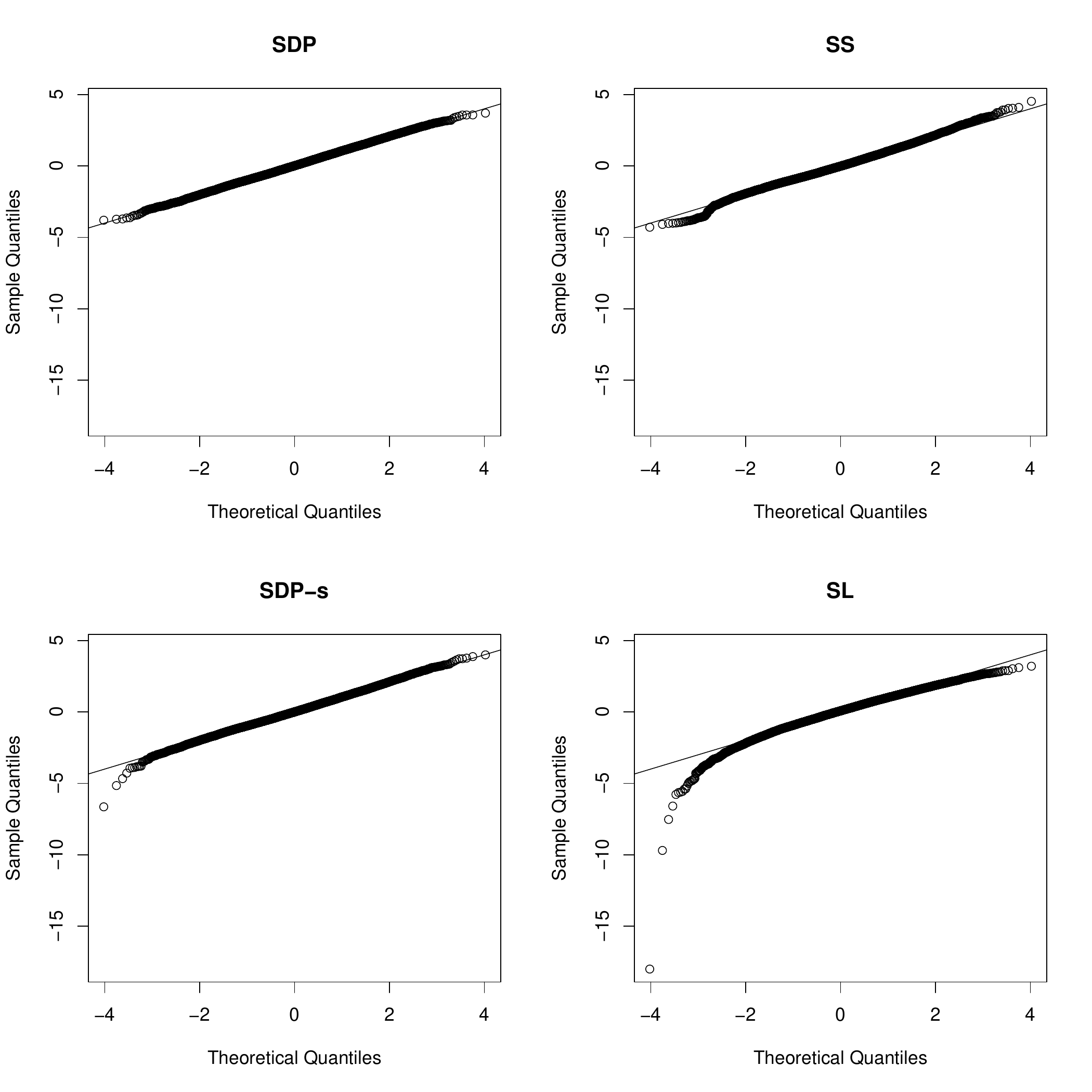}
\caption{QQ-plots of standardized residuals in the four models.}
\label{fig:resid2}
\end{figure}

For comparing the four models, we calculated AIC and BIC given in the end of Section \ref{sec:trans}.
We reported the values in Table \ref{tab:pov-model}, which shows that both AIC and BIC values of the SDP model are significantly smaller than those of the other models.
Moreover, to see the adequacy of normality assumption of $v_i$ and $\ep_{ij}$ in the models (\ref{pv-model}), we computed 
$$
\widehat{v}_i^{\ast}=\frac{\tah}{\sih^2+n_i\tah^2}\sum_{j=1}^{n_i}(\widehat{H}(y_{ij})-\x_{ij}^t\bbeh), \ \ \ \ \ \text{and} \ \ \ \ 
\widehat{\ep}_{ij}^{\ast}=\sih^{-1}\left\{\widehat{H}(y_{ij})-\x_{ij}^t\bbeh-\tah \widehat{v}_i^{\ast}\right\},
$$
and their QQ-plots are shown in Figures \ref{fig:resid1} and \ref{fig:resid2}.
Although the normality of $v_i$ seems plausible in all the four models from Figure \ref{fig:resid1}, the normality of $\ep_{ij}$ in the SL model would not be appropriate from Figure \ref{fig:resid2}.

Finally, we calculated the estimates of poverty indicators based on FGT poverty measures given in (\ref{pr}), where we set $z$ as the $0. 6$ times the median of $E_{ij}$'s.
In particular, we estimated the poverty rate ($\alpha=0$) and poverty gap ($\alpha=1$).
Since the auxiliary variables of non-sampled units in five provinces are available, we computed the estimates of poverty indicators of the provinces based on the four models in (\ref{pv-model}) with 100 Monte Carlo samples as well as the direct estimator (DE).
Moreover, we computed $95\%$ confidence intervals (CI) of the poverty indicators based on 100 bootstrap samples.
The results are given in Table \ref{tab:pov-res}.
It is observed that the four model-based estimates are very different from DE even in the provinces with relatively large number of sampled units (e.g. Sevilla).
Moreover, the SDP and SDP-s models provide quite similar estimates while the SL model provide relatively different estimates from the others.
However, based on AIC and BIC values and QQ-plot in Figure \ref{fig:resid2}, the validity of SL method would be doubtful, so that the estimates given in Table \ref{tab:pov-res} would not be reliable.

\begin{table}[!htb]
\caption{
AIC and BIC of the four models.
\label{tab:pov-model}
}
\begin{center}
\begin{tabular}{ccccccccccccccccccccccccccccc}
\hline
&& SDP & SDP-s & SS & SL\\
\hline
AIC &  & 347778.2 & 347840.0 & 348123.8 & 348870.7 \\
BIC &  & 347886.7 & 347940.8 & 348232.4 & 348963.8 \\
\hline
\end{tabular}
\end{center}
\end{table}

\begin{table}[!htb]
\caption{Estimates and $95\%$ empirical Bayes confidence intervals of poverty rates and poverty gap based on the direct estimator and four model based methods in five provinces.
\label{tab:pov-res}
}
\begin{center}
Poverty rate ($\alpha=0$)
\begin{tabular}{ccccccccccccccccccccccccccccc}
\hline
Province & $n_i$ &  &  & DE & SDP & SDP-s & SS & SL\\
\hline
 &  &  & CI (upper) & - & 23.85 & 24.11 & 23.03 & 25.45\\
Avila  & 58 &  & Estimate & 8.62 & 17.81 & 18.05 & 18.20 & 19.51\\
 &  &  & CI (lower) & - & 12.33 & 12.50 & 14.61 & 14.38\\
\hline
 &  &  & CI (upper) & - & 28.06 & 28.26 & 28.11 & 30.07\\
Tarragona  & 134 &  & Estimate & 29.17 & 25.95 & 26.37 & 26.40 & 28.03\\
 &  &  & CI (lower) & - & 23.71 & 24.35 & 24.32 & 26.16\\
\hline
 &  &  & CI (upper) & - & 34.42 & 36.14 & 35.96 & 38.38\\
Santander  & 434 &  & Estimate & 29.31 & 31.93 & 32.94 & 33.04 & 35.63\\
 &  &  & CI (lower) & - & 29.65 & 30.30 & 30.34 & 32.25\\
\hline
 &  &  & CI (upper) & - & 28.55 & 28.35 & 28.44 & 29.70\\
Sevilla & 482 &  & Estimate & 5.00 & 25.77 & 25.89 & 26.27 & 27.06\\
 &  &  & CI (lower) & - & 23.50 & 23.37 & 24.02 & 24.45\\
\hline
 &  &  & CI (upper) & - & 42.31 & 42.42 & 44.15 & 46.47\\
Oviedo  & 803 &  & Estimate & 33.33 & 37.47 & 37.52 & 37.55 & 40.57\\
 &  &  & CI (lower) & - & 31.35 & 32.60 & 31.30 & 34.18\\
\hline
\end{tabular}

\vspace{0.5cm}
Poverty gap ($\alpha=1$)
\begin{tabular}{ccccccccccccccccccccccccccccc}
\hline
Province & $n_i$ &  &  & DE & SDP & SDP-s & SS & SL\\
 \hline
 &  &  & CI (upper) & - & 8.11 & 8.38 & 8.16 & 7.91\\
Avila  & 58 &  & Estimate & 2.07 & 5.67 & 5.86 & 6.13 & 5.70\\
 &  &  & CI (lower) & - & 3.61 & 3.71 & 4.68 & 3.92\\
 \hline
 &  &  & CI (upper) & - & 10.11 & 10.36 & 10.60 & 9.86\\
Tarragona  & 134 &  & Estimate & 7.69 & 9.14 & 9.47 & 9.80 & 9.01\\
 &  &  & CI (lower) & - & 8.14 & 8.52 & 8.80 & 8.24\\
 \hline
 &  &  & CI (upper) & - & 13.18 & 14.27 & 14.70 & 13.66\\
Santander  & 434 &  & Estimate & 8.63 & 11.96 & 12.63 & 13.09 & 12.35\\
 &  &  & CI (lower) & - & 10.82 & 11.30 & 11.69 & 10.83\\
 \hline
 &  &  & CI (upper) & - & 10.66 & 10.62 & 11.06 & 9.95\\
Sevilla & 482 &  & Estimate & 2.75 & 9.31 & 9.48 & 9.99 & 8.84\\
 &  &  & CI (lower) & - & 8.27 & 8.33 & 8.89 & 7.75\\
 \hline
 &  &  & CI (upper) & - & 17.38 & 17.64 & 19.19 & 17.68\\
Oviedo  & 803 &  & Estimate & 11.66 & 14.76 & 14.94 & 15.43 & 14.62\\
 &  &  & CI (lower) & - & 11.57 & 12.33 & 12.13 & 11.54\\
\hline
\end{tabular}

\end{center}
\end{table}

\section{Conclusions and Discussion}\label{sec:conc}
We have introduced the use of the parametric family of transformations for estimating (predicting) general area-specific parameters based on the mixed effects models.
We have provided the best predictor of the parameter as well as the maximum likelihood method for estimating model parameters.
Moreover, for measuring variability of the predictor, we constructed the mpirical Bayes confidence interval of the area parameter.
The simulation and empirical studies have revealed that the use of parametric transformations would improve the prediction accuracy of the existing method using specified transformations.

As demonstrated in the simulation studies, the proposed method using the parametric family of transformations outperformed the prediction method using specified transformations when the specified transformation is not true.
Hence, the proposed method would be promising and recommended as an alternative tool for prediction methods with specified transformations.
However, when the estimated transformation is close to a well-known one, we may not necessarily employ the proposed method and it would be recommended to simply use the well-known transformation.     
For example, when the estimate of $\lambda$ in the DP transformation is close to $0$, it would be better to simply use the log-transformation.

In this paper, we developed the methodology under the situation where the random effects $v_i$'s are mutually independent, following Molina and Rao (2010). 
However, $v_i$'s might be spatially correlated in some cases and several works have been focused on introducing spatial correlations in small area estimation (e.g. Pratesi and Salvati, 2009; Schmit et al., 2016).
The detailed discussion introducing spatial correlation in this context is left to a future work.

Although we considered an empirical Bayes approach in this paper, the hierarchical Bayes approach as considered in Molina et al. (2014), by assigning some prior distributions for model parameters, would be useful as well.
Moreover, it would be interesting to consider the use of the penalized spline method for modeling the regression part (e.g. Opsomer et al., 2008) or a semiparametric transformation approach (e.g. Nesser, et al., 1996) rather than the full parametric approach, which would achieve more flexible modeling whereas both would be computationally burdensome.  
The detailed investigation are left to valuable future studies.

\ \\
{\bf Acknowledgement} \ \ 
The authors are supported by Grant-in-Aid for Scientific Research (18K12757, 15H01943 and 26330036) from Japan Society for the Promotion of Science.

\vspace{1cm}
\begin{center}
{\bf Appendix}
\end{center}

\vspace{0.2cm}
\noindent
{\bf A1. Proof of Theorem \ref{thm:asymp}.}\ \ \ \ 
From the likelihood function (\ref{ML}), the derivatives are given by 
\begin{align*}
\frac{\partial L}{\partial\bbe}
&=\sum_{i=1}^m\X_i^t\bSi_i^{-1}\z_i, \ \ \ \ 
\frac{\partial L}{\partial\tau^2}
=-\frac12\sum_{i=1}^m\1_{n_i}^t\bSi_i^{-1}\1_{n_i}-\frac12\sum_{i=1}^m\z_i^t\bSi_i^{-1}\1_{n_i}\1_{n_i}^t\bSi_i^{-1}\z_i\\
\frac{\partial L}{\partial\si^2}
&=-\frac12\sum_{i=1}^m\tr(\bSi_i^{-1})-\frac12\sum_{i=1}^m\z_i^t\bSi_i^{-2}\z_i, \\ 
\frac{\partial L}{\partial\la}
&=-\sum_{i=1}^m\z_i^t\bSi_i^{-1}H_{\la}^{(1)}(y_i)+\sum_{i=1}^m\sum_{j=1}^{n_i}\frac{\partial}{\partial\la}\log H_{\la}'(y_{ij}),
\end{align*}
where $\z_i=H_{\la}(y_i)-\X_i\bbe$.
Since $\E[\z_i]=\0$, it follows that $\E[\partial^2L/\partial\bbe \partial\tau^2]=\E[\partial^2L/\partial\bbe \partial\si^2]=\0$.
The other elements of the Fisher information can be obtained by a straightforward calculation.
Moreover, under Assumptions \ref{as:trans} and \ref{as:model}, each element of the Fisher information matrix is finite, so that the asymptotic normality of $\bphih$ follows.


\vspace{0.4cm}
\noindent
{\bf A2. Proof of Theorem \ref{thm:cp1}.}\ \ \ \ 
Let $\bphi_0$ be the true values of parameters.
It suffices to show that $P(\mu_i\leq Q_{a}(y_i,\bphih))=a+O(m^{-1})$ for $a\in (0,1)$.
We first note that 
$$
P(\mu_i\leq Q_{a}(y_i,\bphih))=
\E[P(\mu_i\leq Q_{a}(y_i,\bphih)|\y_s)]=
\E[F(Q_{a}(y_i,\bphih);y_i,\bphi_0)],
$$
where $F(\cdot;y_i,\bphi_0)$ is a distribution function of $\mu_i$ given $y_i$.
Let $G(y_i,\bphih,\bphi_0)=F(Q_{a}(y_i,\bphih);y_i,\bphi_0)$, noting that $0\leq G(y_i,\bphih,\bphi_0)\leq 1$ and $G(y_i,\bphi_0,\bphi_0)=a$. 
The Taylor expansion of $G(y_i,\bphih,\bphi_0)$ shows that 
\begin{align*}
G(y_i,\bphih,\bphi_0)&
=G(y_i,\bphi_0,\bphi_0)+\sum_{j}G_{\phi_j}(y_i,\bphi,\bphi_0)\big|_{\bphi=\bphi_0}(\phih_j-\phi_j)\\
&+\frac12\sum_{j,k}G_{\phi_j\phi_k}(y_i,\bphi,\bphi_0)\big|_{\bphi=\bphi_0}(\phih_j-\phi_j)(\phih_k-\phi_k)\\
&+\frac16\sum_{j,k,\ell}G_{\phi_j\phi_k\phi_\ell}(y_i,\bphi,\bphi_0)\big|_{\bphi=\bphi^{\ast}}(\phih_j-\phi_j)(\phih_k-\phi_k)(\phih_\ell-\phi_\ell),
\end{align*}
where $\bphi^{\ast}$ is on the line connecting $\bphih$ and $\bphi_0$. 
Then, it follows that 
\begin{align*}
P(\mu_i\leq Q_{a}(y_i,\bphih))
&=\E[G(y_i,\bphih,\bphi_0)]
=a+R_1+\frac12R_2+\frac16R_3,
\end{align*}
where 
\begin{align*}
R_1&=\E\left[G_{\bphi}(y_i,\bphi,\bphi_0)\big|_{\bphi=\bphi_0}(\bphih-\bphi_0)\right],\\ 
R_2&=\sum_{j,k}\E\left[G_{\phi_j\phi_k}(y_i,\bphi,\bphi_0)\big|_{\bphi=\bphi_0}(\phih_j-\phi_j)(\phih_k-\phi_k)\right],\\
R_3&=\sum_{j,k,\ell}\E\left[G_{\phi_j\phi_k\phi_\ell}(y_i,\bphi,\bphi_0)\big|_{\bphi=\bphi^{\ast}}(\phih_j-\phi_j)(\phih_k-\phi_k)(\phih_\ell-\phi_\ell)\right].
\end{align*}
Using the Cauchy-Schwarz inequality, we have
\begin{align*}
&\E\left[G_{\phi_j\phi_k}(y_i,\bphi,\bphi_0)\big|_{\bphi=\bphi_0}(\phih_j-\phi_j)(\phih_k-\phi_k)\right]\\
& \hspace{2cm}\leq 
\left\{\E[(\phih_j-\phi_j)^4]\right\}^{\frac14}\left\{\E[(\phih_k-\phi_k)^4]\right\}^{\frac14}\sqrt{\E\left[G_{\phi_j\phi_k}(y_i,\bphi,\bphi_0)^2\big|_{\bphi=\bphi_0}\right]}.
\end{align*}
From the asymptotic normality of $\bphih$ given in Theorem \ref{thm:asymp}, it holds that $\E[|\phih_k-\phi_k|^r]=O(m^{-r/2})$.
Moreover, since $0\leq G(y_i,\bphi,\bphi_0)\leq 1$ and $\bphi_0$ is an interior point, it holds $|G(y_i,\bphi_1,\bphi_0)-G(y_i,\bphi_2,\bphi_0)|\leq 2$ for all $\bphi_1,\bphi_2\in N_{\bphi_0}$ with $N_{\bphi_0}=\{\bphi; \|\bphi-\bphi_0\|\leq \ep\}$, thereby the partial derivatives of $G(y_i,\bphi,\bphi_0)$ at $\bphi=\bphi_0$ are bounded.
Then, we obtain $R_2=O(m^{-1})$.
Using a similar evaluation, we can show that $R_3=O(m^{-1})$.
Regarding $R_1$, it is noted that 
$$
\E\left[G_{\bphi}(y_i,\bphi,\bphi_0)\big|_{\bphi=\bphi_0}(\bphih-\bphi_0)\right]
=
\E\left[G_{\bphi}(y_i,\bphi,\bphi_0)\E[\bphih-\bphi_0|y_i]\right].
$$
From Lohr and Rao (2009), it holds $\E[\bphih-\bphi_0|y_i]=m^{-1}\b_{\bphi}-\I_{\bphi}^{-1}\partial L_i(y_i,\bphi_0)/\partial\bphi+o_p(m^{-1})$, where $\sum_{i=1}^mL_i(y_i,\bphi_0)\equiv L(\bphi)$ and $b_{\bphi}=\lim_{m\to\infty}m\E[\bphih-\bphi_0]$ is the asymptotic bias of $\bphih$.
Hence, we have
\begin{align*}
&\E\left[G_{\bphi}(y_i,\bphi,\bphi_0)\E[\bphih-\bphi_0|y_i]\right]\\
&\ \ \ \ \ \ \ =\frac1m\E\left[G_{\bphi}(y_i,\bphi,\bphi_0)\right]\b_{\bphi}
-\E\left[G_{\bphi}(y_i,\bphi,\bphi_0)\I_{\bphi}^{-1}\frac{\partial}{\partial\bphi}L_i(y_i;\bphi_0)\right]+o(m^{-1}),
\end{align*}
which is $O(m^{-1})$.
Therefore, the proof is completed.


\vspace{0.4cm}
\noindent
{\bf A3. Proof of Theorem \ref{thm:cp2}.}\ \ \ \ 
From the proof of Theorem \ref{thm:cp1}, we have
$$
F_a(\bphi_0)\equiv P(\mu_i\leq Q_{a}(y_i,\bphih))=a+\frac{c(a,\bphi_0)}{m}+o(m^{-1}),
$$
where $c(a,\bphi)$ is a smooth function of $a$ and $\bphi$.
Take $a^{\ast}$ and $\widehat{a}^{\ast}$ so that they satisfy $F_{a^{\ast}}(\bphi_0)=a$ and $F_{\widehat{a}^{\ast}}(\bphih)=a$, respectively. 
Then, it holds $\widehat{a}^{\ast}-a^{\ast}=o_p(1)$ since $\bphih-\bphi=o_p(1)$.
From the above expansion, we have 
$$
\widehat{a}^{\ast}-a^{\ast}=-\frac1m\left\{c(\widehat{a}^{\ast},\bphih)-c(a^{\ast},\bphi_0)\right\}+o(m^{-1}),
$$
so that $\widehat{a}^{\ast}-a^{\ast}=o_p(m^{-1})$.
Hence, it follows that 
$$
P(\mu_i\leq Q_{\widehat{a}^{\ast}}(y_i,\bphih))
=P(\mu_i\leq Q_{a^{\ast}}(y_i,\bphih))+o(m^{-1})
=a+o(m^{-1}),
$$
which completes the proof.


\vspace{0.4cm}
\noindent
{\bf A4. Checking assumptions of transformations.}\ \ \ \ 
We here check the assumption 3 in Assumption \ref{as:trans} for the dual power (DP) transformation (\ref{DPT}) and sinh-arcsinh (SS) transformation (\ref{SS}).

\medskip\noindent
{\bf (DP transformation)} \ \ \ \
We first note that $H_{\la}^{-1}(x)=O(x^{1/\la})$ as $x\to\infty$.
By putting $x=-t$ for $t>0$, we have
$$
H_{\la}^{-1}(x)=(\sqrt{1+\la^2t^2}-\la t)^{1/\la}=\frac{1}{(\sqrt{1+\la^2t^2}+\la t)^{1/\la}}=O(t^{-1/\la})
$$
as $t\to\infty$.
A straightforward calculation shows that 
$$
\frac{\partial H_{\la}(x)}{\partial\la}=\frac{x^{\la}\log x+x^{-\la}\log x}{2\la}+\frac{x^{\la}-x^{-\la}}{2\la^2},
$$
thereby, it follows that 
$$
\bigg|\frac{\partial H_{\la}}{\partial\la}(H_{\la}^{-1}(x))\bigg|=O(|x|\log |x|)+O(|x|^{-1}\log |x|)+O(|x|)+O(|x|^{-1})=O(|x|\log |x|)
$$
as $|x|\to\infty$.
Moreover, since 
$$
\frac{\partial^2H_{\la}(x)}{\partial\la^2}=\frac{x^{\la}(\log x)^2-x^{-\la}(\log x)^2}{2\la}-\frac{x^{\la}-x^{-\la}}{\la^3},
$$
a similar evaluation leads to 
$\big|\partial^2H_{\la}(w)/\partial\la^2\big|=O(|x|(\log|x|)^2)$
as $|x|\to\infty$.
Regarding $\partial^2\log H_{\la}'(x)/\partial\la^2$, it holds that 
$$
\bigg|\frac{\partial^2\log H_{\la}'(w)}{\partial\la^2}\bigg|=\bigg|\frac{4(\log w)^2}{w^2(w^{\la-1}+w^{-\la-1})^2}\bigg|=O((\log|x|)^2|x|^2)
$$
as $|x|\to\infty$, so that the DP transformation satisfies the assumption.
When the location parameter is used, namely, $H_{\la,c}(x)=\{(x+c)^{\la}-(x+c)^{-\la}\}/2\la$, it is noted that $\partial^k H_{\la,c}(x)/\partial c^k=\partial^k H_{\la,c}(x)/\partial x^k$, so that the quite similar evaluation shows that the shifted-DP transformation also satisfies the assumption.

\medskip\noindent
{\bf (SS transformation)} \ \ \ \
It follows that 
$$
\frac{\partial H_{a,b}(x)}{\partial a}=-\cosh(b\sinh^{-1}(x)-a), \ \ \ \ 
\frac{\partial H_{a,b}(x)}{\partial b}=\cosh(b\sinh^{-1}(x)-a)\sinh^{-1}(x).
$$
Note that $\sinh^{-1}(x)=O(\log|x|)$ as $|x|\to\infty$, so that $H_{a,b}^{-1}(x)=O(\exp(b^{-1}\log|x|))=O(|x|^{1/b})$.
Then, we have
\begin{align*}
&\frac{\partial H_{a,b}}{\partial a}(H_{a,b}^{-1}(x))=O(\exp(b\log|x|^{1/b}))=O(|x|),\\ 
&\frac{\partial H_{a,b}}{\partial b}(H_{a,b}^{-1}(x))=O(\exp(b\log|x|^{1/b})\log|x|^{1/b})=O(|x|\log|x|),
\end{align*}
as $|x|\to\infty$.
Moreover, it holds that 
\begin{align*}
&\frac{\partial^2 H_{a,b}(x)}{\partial^2 a}=\sinh(b\sinh^{-1}(x)-a), \ \ \ \ 
\frac{\partial^2 H_{a,b}(x)}{\partial^2 b}=\sinh(b\sinh^{-1}(x)-a)\{\sinh^{-1}(x)\}^2\\
&\frac{\partial^2 H_{a,b}(x)}{\partial a \partial b}=-\sinh(b\sinh^{-1}(x)-a)\sinh^{-1}(x),
\end{align*}
thereby a similar evaluation shows that $\partial^2 H_{a,b}(x)/\partial^2 a=O(|x|)$, $\partial^2 H_{a,b}(x)/\partial^2 b=O(|x|(\log|x|)^2)$ and $\partial^2 H_{a,b}(x)/\partial a\partial b=O(|x|\log|x|)$ as $|x|\to\infty$.
On the other hand, a straightforward calculation shows that 
\begin{align*}
\frac{\partial}{\partial a}\log H_{a,b}'(x)=\frac{H_{a,b}(x)}{1+H_{a,b}(x)^2}\frac{\partial H_{a,b}(x)}{\partial a}, \ \ \ \ 
\frac{\partial}{\partial b}\log H_{a,b}'(x)=\frac1b+\frac{H_{a,b}(x)}{1+H_{a,b}(x)^2}\frac{\partial H_{a,b}(x)}{\partial b},
\end{align*}
which are bounded by the function $\partial H_{a,b}(x)/\partial a$ and $\partial H_{a,b}(x)/\partial b$, respectively.
It is not difficult to show that the second partial derivatives of $\log H_{a,b}'(x)$ are bounded by polynomial functions of the second partial derivatives of $H_{a,b}(x)$ and $H_{a,b}(x)$, thereby the assumption is satisfied.


\end{document}